\documentclass{emulateapj}
\usepackage{graphicx}
\usepackage{amssymb}
\usepackage{amsmath}
\shorttitle{Gravitational Infall onto Filaments II}
\shortauthors{Heitsch}

\newcommand{\f}{\frac}
\begin{document}
\title{Gravitational Infall onto Molecular Filaments II. Externally Pressurized Cylinders} 

\author{Fabian Heitsch\altaffilmark{1}}
\altaffiltext{1}{Department of Physics and Astronomy, University of North Carolina Chapel Hill, Chapel Hill, NC 27599-3255}
\email{fheitsch@unc.edu}

\begin{abstract}
In an extension of Fischera \& Martin (2012a) and Heitsch (2013), two aspects of the evolution of externally pressurized, hydrostatic filaments 
are discussed. (a) The free-fall accretion of gas onto such a filament will lead to filament parameters (specifically, FWHM--column density
relations)
inconsistent with the observations of Arzoumanian et al. (2011), except for two cases: For low-mass, isothermal filaments, agreement
is found as in the analysis by Fischera \& Martin (2012b). Magnetized cases, for which the field scales weakly with the density as $B\propto n^{1/2}$,
also reproduce observed parameters. 
(b) Realistically, the filaments will be embedded not only in gas of non-zero pressure, but also of non-zero density. Thus, the appearance
of sheet-embedded filaments is explored. Generating a grid of filament models and comparing the resulting column density ratios and
profile shapes with observations suggests that the three-dimensional filament profiles are intrinsically flatter than 
isothermal, beyond projection and evolution effects. 
\end{abstract}


%%%%%%%%%%%%%%%%%%%%%%%%%%%%%%%%%%%%%%%%%%%%%%%%%%%%%%%%%%%%%%
\section{Motivation}
In a previous study I explored the role of accretion for the evolution of an idealized filamentary
molecular cloud \citep[][H13]{2013ApJ...769..115H}. For consistency with observed filament parameters, specifically with the discussion
by \citet{2011A&A...529L...6A}, the radial density profile was explicitly set, via
\begin{equation}
  \rho(R) = \rho_c\left(1+\left(\frac{R}{R_0}\right)\right)^{-p/2}\label{e:rhoRofp},
\end{equation}
with the core radius $R_0$ and the central density $\rho_c$. 
Herschel studies of filaments \citep{2011A&A...529L...6A} determine the value of the exponent $p$ to be clearly less than the isothermal $p=4$
\citep{1964ApJ...140.1056O}, though \citet{2011A&A...533A..34H} find steeper exponents consistent with $p=4$ in some cases for filaments in Taurus.
Flatter than isothermal profiles have been ascribed to magnetic fields \citep{2000MNRAS.311...85F},
accretion, and non-isothermality \citep{1999ApJ...515..239N}.
H13 was mainly interested in the effect of accretion on a filament's evolution, but less in its structure. Yet, a
more physical way to set $p$ would be desirable.

\citet[][FM12a]{2012A&A...542A..77F} study the structure of infinite, externally pressurized cylinders and discuss how the filament
profiles depend on the external pressure. They argue that for infinite overpressure (or, for the cylinder in a vacuum), the
isothermal solution is recovered, while for finite pressures the profiles flatten, reaching lower $p$. 
They apply their models to a sample of four filaments \citep[][FM12b]{2012A&A...547A..86F}, finding good agreement with their
isothermal cylinder model.

My goal here is to elucidate how the evolution of externally pressurized, isothermal cylinders depends on accretion, thus combining
the analysis of FM12a and H13 (\S\ref{s:accretion}, \ref{s:discussion}). I will discuss isothermal, turbulent and magnetic cases.
It also seems reasonable to explore the effect of ram pressure on the filament structure and evolution, since 
the infalling gas is expected to exert an additional pressure on the filament. 
I conclude with a tool to estimate the evolutionary stage of filaments embedded in flattened clouds (or sheets) in \S\ref{s:sheetfilaments}.

%%%%%%%%%%%%%%%%%%%%%%%%%%%%%%%%%%%%%%%%%%%%%%%%%%%%%%%%%%%%%%
\section{Evolution of an accreting, pressurized filament}\label{s:accretion}
The filament is modeled as an infinite, externally pressurized, and isothermal cylinder, accreting gas at free-fall velocities. Though
free-fall accretion is certainly an extreme assumption \citep[e.g.][]{2012A&A...540A.104M}, I intend it as a counter-point to the
more classical equilibrium consideration.

\subsection{Basic equations}
The goal is to calculate the time evolution of the line mass $m$ due to accretion. For hydrostatic
isothermal cylinders, radially stable solutions can only be found for 
\begin{equation}
  m<m_{cr}\equiv \frac{2c_s^2}{G}=16.3\left(\f{T}{10\mbox{K}}\right)\,\mbox{M}_\odot\mbox{ pc}^{-1}
  \label{e:mlinecrit}
\end{equation}
\citep[e.g.][]{1964ApJ...140.1056O}. Following FM12a, the line
mass is given in terms of a criticality parameter,
\begin{equation}
  f\equiv \frac{m}{m_{cr}} = \frac{mG}{2c_s^2} < 1,
  \label{e:fcrit}
\end{equation}
where $c_s$ is the isothermal sound speed.
The core radius $R_0$ of equation~\ref{e:rhoRofp} is
set to the isothermal value,
\begin{equation}
  R_0^2 = \f{m_{cr}}{\pi \rho_c}.
  \label{e:R0}
\end{equation}

The mass accretion onto the filament can be described by
\begin{eqnarray}
  \f{dm}{dt}&=&2\pi R_f\rho_{ext} v_R\nonumber\\
            &=&4\pi\rho_{ext} R_f\left(G m(t)\,\ln\left(\f{R_{ref}}{R_f}\right)\right)^{1/2},
  \label{e:dmdt}
\end{eqnarray}
where the steady-state, free-fall velocity around an infinite cylinder
is given by 
\begin{equation}
  v_R = 2\left(Gm\,\ln\f{R_{ref}}{R}\right)^{1/2}.
  \label{e:vrad}
\end{equation}
\citep[see][]{2009ApJ...704.1735H}. In equation~\ref{e:dmdt}, $R_f$ is the filament radius,
and $R_{ref}$ is a reference radius (actually, the integration constant), from which the
fluid parcels start their travel to the filament (see discussion in H13, and \S\ref{sss:refradius}). 
Thus, the accretion rate
will depend on the evolutionary stage of the filament via the line mass $m(t)$, and
on the filament radius $R_f$. The ambient density $\rho_{ext}$ is also a free parameter (see \S\ref{ss:BCs}).

\subsection{Boundary Conditions and Profiles}\label{ss:BCs}
To determine $R_f$ and associated quantities, boundary conditions for the filament are
needed. As do FM12a, I assume that the filament 
is embedded in a background medium of pressure $p_{ext}$, with the filament 
in pressure balance such that $p(R_f)=p_{ext}$.  To simplify
the discussion, isothermality is assumed beyond the cylinder, and hence $\rho_{ext} = \rho(R_f)$.

\begin{figure*}
  \includegraphics[width=\textwidth]{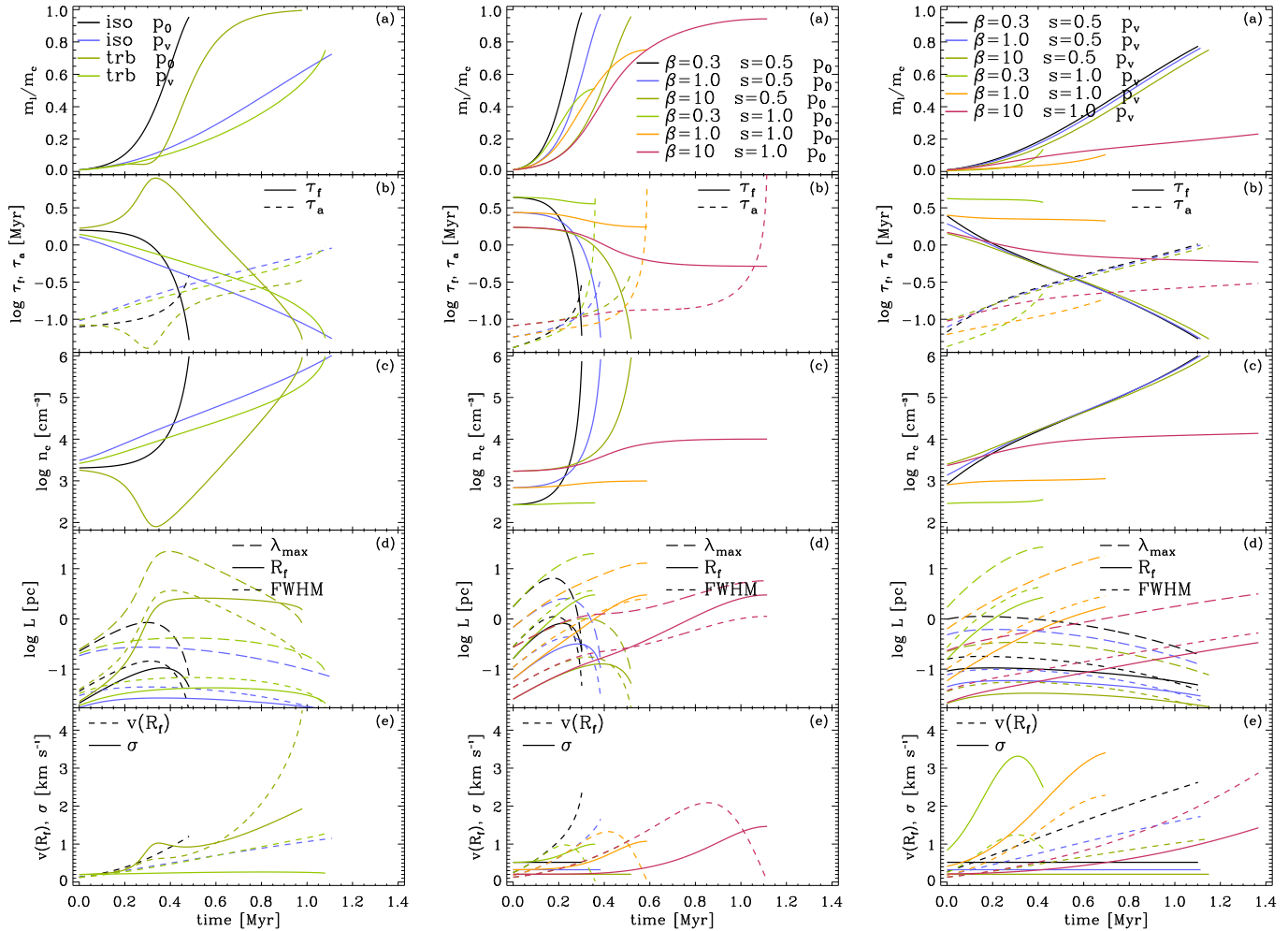}
  \caption{\label{f:plotsingle_all} Time evolution of filament parameters, for all cases considered (see text). From left to right:
           All isothermal and turbulent cases with constant and varying external pressure, all magnetic cases at constant pressure,
           all magnetic cases with varying external pressure. Each column has the following panels:
           (a) Line mass over critical value (eq.~\ref{e:mlinecrit}). (b) Fragmentation
           timescale $\tau_f$ (solid line) and accretion timescale $\tau_a$ (dashed line).
           (c) Central filament (atomic) density $n_c$ (eq.~\ref{e:rhoc}). (d) The filament radius $R_f$ (eq.~\ref{e:radmain}) and
           the FWHM in parsec, and the wavelength of the most
           unstable mode (see text). (e) Accretion velocity $v(R_f)$ (dashed line) and modified sound speed (solid line)
           $\sigma$ driven by accretion or magnetic fields.}
\end{figure*}

A brief summary of the results of FM12a can be 
found in Appendix~\ref{a:fischera}, including the expression for the filament radius
\begin{equation}
  R_f = \sigma^2\left(\frac{2f(1-f)}{\pi G p_{ext}}\right)^{1/2}
  \label{e:radmain}
\end{equation}
needed to evaluate equation~\ref{e:dmdt}. Here, $\sigma$ is a proxy for the effective sound speed.
The goal is to consider a wider variety of physical environments, exploring their effect on 
the accretion and on observable filament parameters.
Figure~\ref{f:plotsingle_all} summarizes the relevant physical quantities for all cases considered.

Before discussing each case in turn, this is as good a moment as any to point out differences in some
of the diagnostic quantities with respect to H13. There, I compared the timescales for accretion 
and gravitational fragmentation with the result that accretion occurs on similar timescales
as fragmentation, and thus should not be neglected when discussing the evolution of filaments. The
expressions for the gravitational fragmentation time scale $\tau_f$ and maximum growth length scale $\lambda_{max}$ 
were taken from \citet{1995ApJ...438..226T}, who in turn used the expressions of \citet{1987PThPh..77..635N},
\begin{eqnarray}
  \tau_f&=& \frac{2.95}{\sqrt{4\pi G \rho_c}}\\
  \lambda_{max}&=&\frac{22.1c_s}{\sqrt{4\pi G\rho_c}}.
\end{eqnarray}
Yet, these expressions assume that the ambient pressure is zero. This assumption is no longer valid.
FM12a calculated polynomial fits for $\tau_f$ and $\lambda_{max}$ (see their appendix E, equation~E.1 and table~E.1.). 
Note that I will use the same expressions
for the magnetic cases, based on the argument that for axial magnetic fields, the longitudinal gravitational
instability will be nearly unaffected, while the growth rates for varicose instabilities are substantially higher,
and thus not of interest in our case (FM12a).

\subsubsection{The Isothermal Case}\label{sss:iso0}
Equations~\ref{e:fcrit}, \ref{e:dmdt}, and \ref{e:radmain} result in an
ordinary differential equation that can be integrated with a 4th order Runge-Kutta method\footnote{Were it not
for the logarithmic term, the ODE could be integrated directly.}. The
RHS only depends on the line mass $m$. The left column of Figure~\ref{f:plotsingle_all} shows the results
(black lines). For consistency with FM12a, I use $p_{ext}=2\times10^4$~K~cm$^{-3}$,
and $T=10$~K, yielding identical results. Specifically, the FWHM peaks at $0.14$~pc. The FWHM is calculated 
numerically for the position $x=R_{FWHM}/R_f$ in the column density profile $N(0)/2 = N(x)$, 
combining equations~15 and 18 of FM12a, with FWHM$=2R_{FWHM}$.

\subsubsection{Effects of Ram Pressure}\label{sss:iso1}
Since the filament is accreting mass, the infall could be thought to exert a ram pressure 
\begin{equation}
  p_{ram} = \rho_{ext} v_R^2 = 4\rho_{ext}\,G\,m\,\ln(R_{ref}/R_f)
  \label{e:pram}
\end{equation}
on the filament surface. Thus, the total external pressure is 
\begin{equation}
  p_{ext}=p_{ram}+p_{ext,0}
  \label{e:pext}
\end{equation}
where $p_{ext,0}$ is
the external thermal pressure. The accretion velocity $v_R$ for the isothermal case (left column, black lines)
already suggests that the ram pressure could get substantially larger than the thermal pressure.
Thus, the filament is ``squeezed'', or, equivalently, truncated at smaller $R_f$, leading to an effective
flattening of the profile, since $R_f<R_0$, see also Fig. 3 of FM12a.
The radius $R_f$ now depends on the accretion velocity,
and hence, in a non-trivial manner on itself. The numerical procedure is explained in Appendix~\ref{aa:iso1}.

Blue lines in the left column of Figure~\ref{f:plotsingle_all} summarize the results. 
For the isothermal case with constant external pressure (\S\ref{sss:iso0}), the accretion rate (eq.~\ref{e:dmdt})
increases with increasing filament radius $R_f$ and line mass $m$. Increasing the pressure (in our case by tracking
the ram pressure $\rho_{ext}v_R^2$) reduces the filament radius, and thus the accretion rate. Thus, once 
$p_{ram} \approx p_{ext,0}$, the accretion rate drops below that of the constant pressure model, and thus the filament
growth slows down. This can be seen in Figure~\ref{f:plotsingle_all}(a), left column. The central densities 
end up growing faster initially, due to the overall compression, but eventually, the increasing overpressure leads
to a flattening of the profile (Fig.~3 of FM12a), and thus to lower central densities compared to the constant pressure case.
Comparing the fragmentation and accretion timescales (Fig.~\ref{f:plotsingle_all}(b), left column), we notice that the increasing external pressure
is driving the filament to fragmentation at earlier times.

\subsubsection{Accretion-Driven Turbulence}\label{sss:accturb}
\citet{2010A&A...520A..17K} argue that in many astrophysical objects turbulence is driven
by accretion. For molecular clouds, this point has been made based on simulations
of flow-driven cloud formation
\citep[][see also \citet{2008MNRAS.385..181F} for a more systematic approach]{2007ApJ...657..870V,2008ApJ...674..316H}.
In this scenario, turbulence is a consequence of the formation process
initially, while at later stages, global gravitational accelerations drive ``turbulent'' motions.

\citet{2010A&A...520A..17K} estimate the level of turbulence driven by accretion (their eqs. 2, 3, and 23).
For the purposes here, the characteristic length scale is $2R_f$, the
accretion velocity $v(R_f)$, and the driving efficiency $\epsilon=0.1$ (see their eq. (23)).
Then, the total "sound speed" is given by
\begin{equation}
  \sigma^2=\left(2\epsilon R_f v^2(R_f)\f{dm/dt}{m(t)}\right)^{2/3} + c_s^2.
  \label{e:sigmaturb}
\end{equation}
The choice of $\epsilon=0.1$ errs on the generous side -- \citet{2010A&A...520A..17K} quote
values of a few percent. 
Replacing the sound speed with the velocity dispersion renders the filament radius dependent on $\sigma$ and thus on itself,
hence, the solution needs to be found numerically (see Appendices~\ref{aa:iso2} and \ref{aa:iso3}). 

The results are summarized by the dark and light green lines in the left column of Figure~\ref{f:plotsingle_all}.
The internal motions driven by accretion increase the filament radius drastically -- for the case with constant external pressure by
nearly a factor of $10$ with respect to the isothermal case, consistent with the ratio $\sigma^2/c_s^2$. This case also shows a
peculiar drop in the central density: Due to the higher internal pressure, the core radius $R_0$ increases, thus reducing the central
density. Once sufficient mass has been accreted, the filament contracts further, and the central density increases.
If the external pressure increases, the filament is being compressed again, and the evolution is similar to that of the isothermal
case including accretion pressure, albeit the radii are larger due to the higher internal pressure.  

Summarizing, accretion-driven turbulence reduces the growth rates in all
quantities, but it does not eventually stabilize the filament: the fragmentation timescale eventually wins
over the accretion timescale, consistent with the results of H13. 
Replacing the sound speed by the turbulent velocity $\sigma$ in $\lambda_{max}$
would increase the fragmentation time scale at a given time, but this seems an improper thing to do.

\subsubsection{Accretion of Magnetic Fields}
\citet{2000MNRAS.311...85F,2000MNRAS.311..105F} discussed in great detail equilibrium configurations
of cylinders with a variety of magnetic field geometries. Here, we are interested less in {\em stable}
configurations, but to what extent magnetic fields affect the {\em growth} of the filament.

The effect of magnetic fields can be approximated by modifying the sound speed
$c_s$ in equation~\ref{e:radmain} to
\begin{equation}
  \sigma^2\equiv c_s^2\left(1+\f{2}{\beta_0}\left(\f{n_c}{n_{c0}}\right)^{2s-1}\right),
  \label{e:magsound}
\end{equation}
with the initial plasma parameter
\begin{equation}
  \beta_0 \equiv\f{2c_s^2}{c_A^2}=\f{8\pi c_s^2\rho_{c0}}{B_0^2}.
\end{equation}
Here, $n_c$ is the filament's central density, with the initial condition $\rho_{c0} = \mu m_H n_{c0}$ at $t=0$.
Appendices~\ref{aa:iso4}--\ref{aa:iso5b} provide more details on how to solve the equations.
Because of flux-freezing, a scaling of the magnetic field strength with
density is assumed,
\begin{equation}
  B\propto n^s,
  \label{e:mageos}
\end{equation}
which will depend on the field geometry through the exponent $s$. For fields along the axis of the filament
and for toroidal fields, $s=1$, and for 
%\begin{equation}
%  \sigma^2 = c_s^2\left(1+\f{2}{\beta_0}\f{n_c}{n_{c0}}\right).
%  \label{e:magsound1/2}
%\end{equation}
a uniform field perpendicular to the filament axis, 
$s=1/2$ under mass and flux conservation.
%, resulting in
%\begin{equation}
%  \sigma^2 = c_s^2\left(1+\f{2}{\beta_0}\right).
%  \label{e:magsound1}
%\end{equation}
In the latter case, the magnetic pressure will stay constant, at its initial level. 

The results are summarized in the center and right column of Figure~\ref{f:plotsingle_all}. The same quantities are
shown as for the hydrodynamical case (left column), for three magnetization strengths, $\beta_0=0.3,1.0,10.0$, and
for $s=0.5,1.0$ (eq.~\ref{e:mageos}). Two points are noteworthy: (1) For constant external pressure, a 
strong scaling of the field with density ($s=1$, i.e. toroidal or poloidal fields) effectively shuts down accretion. Both
the line mass and the central density converge to a saturation value set by magnetostatic equilibrium. In these cases,
the accretion velocity drops to zero. For the weak magnetic scaling ($s=1/2$), accretion cannot be stopped -- the system
essentially behaves isothermally, with an increased soundspeed. (2) If the filament is pressurized by ram pressure due
to accretion (right column), the line masses grow for all cases except for $\beta_0=0.3,s=1$.

As for the turbulent case, the fragmentation timescale depends on the field strength through the central
density $n_c$. In difference to the turbulent case, the length scale of maximum growth now depends
directly on the modified sound speed -- consistent with the notion that magnetic fields can suppress gravitational
fragmentation.

While magnetic fields can affect mass accretion onto the filament, they do not prevent it (at least not for
reasonable choices of magnetization). Fragmentation wins over accretion once the filament becomes critical.
I forego the discussion of turbulence in combination with magnetic fields. There is numerical and analytical evidence
that turbulence combined with magnetic flux loss mechanisms (ambipolar drift and reconnection) efficiently
reduce the dynamical importance of magnetic fields
\citep{1999ApJ...517..700L,2010ApJ...714..442S,2002ApJ...578L.113K,2002ApJ...567..962Z,2004ApJ...603..165H}.

%%%%%%%%%%%%%%%%%%%%%%%%%%%%%%%%%%%%%%%%%%%%%%%%%%%%%%%%%%%%%%
\section{Discussion}\label{s:discussion}

%The striations and "spikes" seen off the Taurus main filaments (e.g. REF Heyer, Tobin) are similarly suggestive of 
%active accretion. The fact that dust polarimetry seems to suggest that the magnetic field is aligned with those
%striations, however, poses some challenges to the overall field geometry, as long as one assumes that the 
%Taurus filaments are not embedded in a sheet (with the field in the plane). 
%This is equally correct for clouds like Taurus, or the Pipe Nebula, 
%as well as for massive IRDCs. 

Before we discuss the results in terms of observations (\S\ref{ss:filevol}), a closer look 
at the assumptions made in the models, and their effects, seems in place (\S\ref{ss:unphyscase}). This will help to rule
out unphysical cases.

\subsection{Model Assumptions and Consequences}\label{ss:unphyscase}

\subsubsection{Hydrostatic Equilibrium}\label{sss:hydrostat}

Models with filament radii approaching $1$~pc (see Fig.~\ref{f:plotsingle_all}) also tend to have large line masses, and thus develop substantial accretion
velocities (eq.~\ref{e:vrad}). For example, the
case "trb p$_0$" shows a dramatic increase in the filament radius around $0.4$~Myr to $\sim2$~pc, with an increase of the 
accretion velocity beyond $2$~km~s$^{-1}$. This violates two assumptions made: the assumption of hydrostatic equilibrium,
and, possibly, the assumption of isothermality (\S\ref{sss:isothermal}).

\begin{figure*}
  \includegraphics[width=\textwidth]{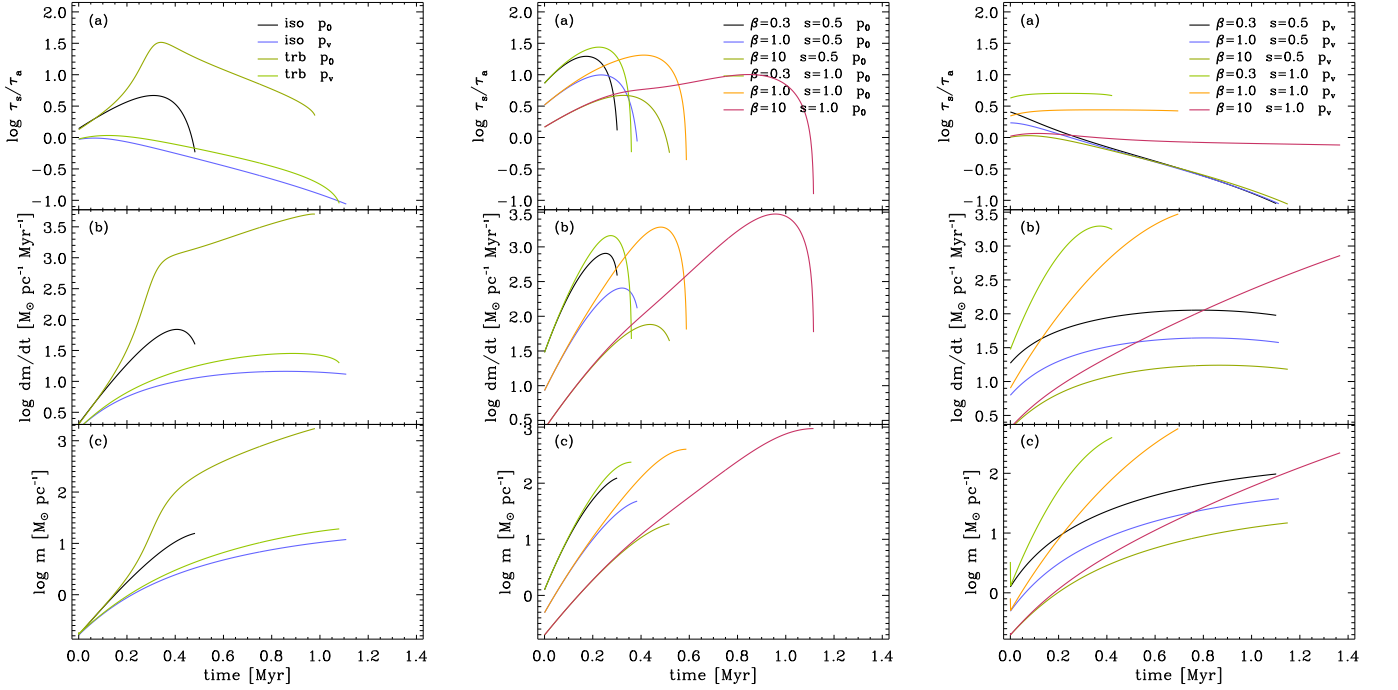}
  \caption{\label{f:plotsingle_times}
           {\em Top row (a):} Logarithm of the ratio between the sound crossing time $\tau_s$ and the accretion time scale $\tau_a$ 
           against time for all models (from left to right:
           isothermal and turbulent, magnetic with constant pressure, and magnetic with varying external pressure, as indicated).
           For $\log \tau_s/\tau_a>0$, the assumption of hydrostatic equilibrium is invalid.
           {\em Center row (b):} Logarithm of the line mass accretion rates against time for all models, as above,
           used to highlight the effects of the filament evolution on the accretion rate.
           {\em Bottom row (c):} Logarithm of line mass against time for all models. Note that the critical line masses (eq.~\ref{e:mlinecrit})
           depend on the effective sound speed, and are generally larger than the isothermal value.}
\end{figure*}

The filament is modeled as a hydrostatic
cylinder, thus, any change of the external pressure needs to be communicated fast enough throughout the cylinder to allow
the internal pressure to adjust. In other words, the ratio of the signal crossing time $\tau_s$ (which, in the isothermal case, is set by the sound speed,
and in the turbulent case is determined by the turbulent {\em rms} velocity $\sigma$) and the accretion time scale $\tau_a=m/\dot{m}$
should be 
\begin{equation}
  \tau_s/\tau_a<1\label{e:hydrostat}
\end{equation}
for hydrostatic equilibrium. Figure~\ref{f:plotsingle_times} shows this ratio for all models, in the same color styles as 
in Figure~\ref{f:plotsingle_all}. For the unmagnetized models, only those with varying external pressure ("iso p$_v$" and "trb p$_v$")
turn out to meet condition~\ref{e:hydrostat}. If $\tau_s/\tau_a>1$, the internal pressure cannot adjust fast enough to keep 
hydrostatic equilibrium, and the cylinder will start to collapse radially. In that case, while the derived line masses are still useable,
other filament parameters such as radii or central densities will be incorrect.

\subsubsection{Reference Radius $R_{ref}$}\label{sss:refradius}
\begin{figure}
  \includegraphics[width=\columnwidth]{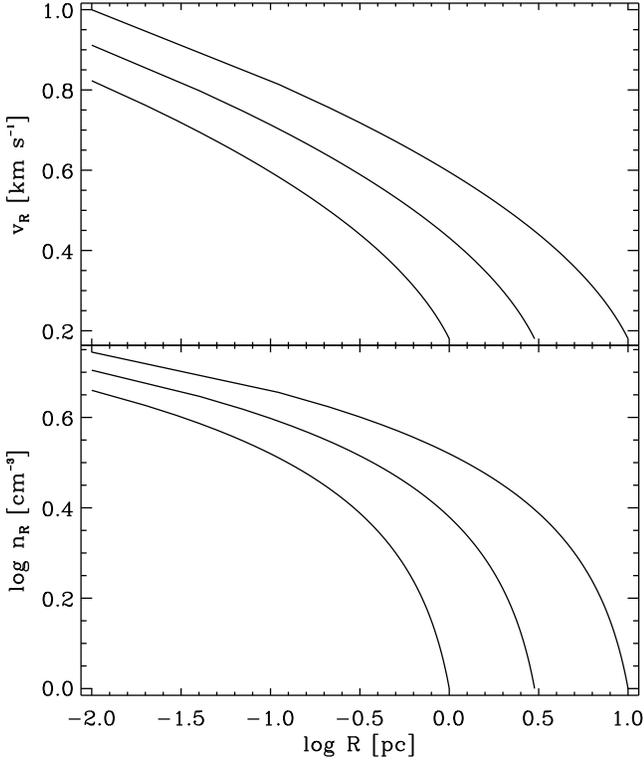}
  \caption{\label{f:vnprofiles} Accretion velocity $v_R$ ({\em top}, eq.~\ref{e:vrad}) against radius, and corresponding over-density profile
           ({\em bottom}) for three reference radii $R_{ref}=1,3,10$~pc, 
           assuming a line mass of $m=16.3$~M$_\odot$~pc$^{-1}$. The velocities stay below $1$~km~s$^{-1}$, and the densities increase
           at most by a factor of $\lesssim 6$.}
\end{figure}

The reference radius $R_{ref}$ needed to calculate the steady-state, free-fall radial velocity profile (eq.~\ref{e:vrad}) is a free parameter
of the model. It has been set to $10$~pc for the current models (compared to $2$~pc in H13). Figure~\ref{f:vnprofiles} (top panel) 
demonstrates the effect of the choice of $R_{ref}$ on $v_R$, for $R_{ref}=1,3,10$~pc. For realistic filament radii of $\sim 0.1$~pc, $v_R$
varies between $0.6$ and $0.85$~km~s$^{-1}$, resulting in a variation in the accretion timescales of $30$\%. These values are consistent with
the accretion velocities derived by \citet{2013ApJ...766..115K} for Serpens South.

\subsubsection{Isothermality}\label{sss:isothermal}
The assumption of isothermality enters in two places. (1) The filament profiles are derived for hydrostatic, infinite, isothermal 
cylinders \citep{1964ApJ...140.1056O}.
\citet{2011A&A...529L...6A} find radial temperature variations strong enough to susceptibly flatten the radial
density profiles of corresponding hydrostatic cylinders, while \citet{2012A&A...547A..86F} rely on dust temperatures derived at
longer wavelengths, where temperature variations are less pronounced. 
Here, I follow the latter authors in assuming a constant temperature, mainly for the purpose to restrict the 
number of free parameters in the model. A more accurate treatment, also more consistent with observational evidence, will be discussed
elsewhere. 

(2) Taken at face value, the accretion velocities derived
from the integration of eq.~\ref{e:dmdt} would lead to shocks at Mach numbers of $10$ or more in some cases, or of shock velocities
of up to a few km~s$^{-1}$ in extreme cases (Fig.~\ref{f:plotsingle_all}). Figure~\ref{f:vnprofiles} (top panel) shows that the steady-state, free-fall velocity
profile (eq.~\ref{e:vrad}) expected for a line mass of $m=16.3$~M$_\odot$~pc$^{-1}$ is generally less than $1$~km~s$^{-1}$, even for exceedingly large
reference radii $R_{ref}$. Assuming $T=10$K, these $v_R$ correspond to Mach numbers of up to $5$. It also should be noted that $v_R\propto \sqrt{m}$, thus
higher line masses do not affect $v_R$ strongly.  
\citet{1983ApJ...264..485D} give peak temperatures of a few hundred Kelvin for
shocks in dense molecular gas ($n=10^4$~cm$^{-3}$) at a shock velocity of $5$~km~s$^{-1}$, and with a radiative shock width of a
$10^{-3}\cdots10^{-2}$~pc. Due to the high densities (and thus, the high cooling rates), the shocks can be treated as isothermal
beyond this length scale, thus justifying the isothermal assumption even for the most strongly accreting filaments in our models. 
The very assumption of an accretion shock may be unrealistic specifically in the cases with extreme $v_R$, since those correspond to
models including accretion-driven turbulence. The turbulence itself implies that a well-defined filament boundary does not exist. 

\subsubsection{Background Density}
The choice of the background
density $n_{ext}$ plays a crucial role for the timescales discussed in \S\ref{sss:hydrostat}: 
the accretion timescale depends linearly on $n_{ext}$ (see eq.~\ref{e:dmdt}),
and our choice of $n_{ext} = P_{ext}/(\mu m_H c_s^2)=2\times10^3$~cm$^{-3}$ seems rather high for a molecular cloud envelope. Choosing
a lower $n_{ext}$ by assuming a higher ambient temperature will extend the timescales by the same factor, thus bringing the
isothermal case with constant pressure ("iso p$_0$") into the regime $\tau_s/\tau_a<1$. 

The background density $n_{ext}$ is assumed to stay constant over the evolution of the filament. This simplification serves mainly for
consistency with previous models (H13), and it is also motivated to some extent by the column density profile
shown in Figure~4 of \citet{2011A&A...529L...6A}. Under more general conditions, the background density may change both in space
(even at $t=0$) and with time, while the filament is accreting. Since it generally will drop, the evolution timescales derived here
are lower limits. Figure~\ref{f:vnprofiles} (bottom) shows the expected free-fall, steady-state density profile corresponding to the
accretion velocities (top panel). The density profiles were found by integrating the continuity equation using eq.~\ref{e:vrad}, and
assuming a non-zero inward velocity component $v_0=c_s$ at $R_{ref}$. 
%Then, 
%\begin{equation}
%  \frac{dn_R}{dR}=-\frac{n_R}{v_R}\frac{dv_R}{dR}-\frac{n_R}{R}
%\end{equation}
%leads to
%\begin{equation}
%  n_R = n_{ext}\frac{v_R}{v_0}, 
%  \label{e:nrad}
%\end{equation}
%with a constant background density at $R=R_{ref}$ of $n_{ext}$. 
For plotting purposes, I set $n_{ext}=1$~cm$^{-3}$ in Fig.~\ref{f:vnprofiles},
yet, since a pressure-less accretion flow is assumed, this choice does not affect the density increase. In any case, the densities vary by
a factor $\lesssim 6$ at most (for $m=16.3$~M$_\odot$~pc$^{-1}$). 

\subsubsection{Mass Reservoir}\label{sss:massreservoir}
The center and bottom row of Figure~\ref{f:plotsingle_times} highlight the mass history of an accreting filament. The center row shows
the mass accretion rates for all models, and the bottom row the actual line masses. Note that the critical line mass changes depending
on the model assumptions, since the effective sound speed $\sigma$ may replace the actual sound speed $c_s$. Thus, all line masses shown
in Figure~\ref{f:plotsingle_times} are subcritical with respect to $\sigma$, but may be supercritical with respect to $c_s$. 

\citet{2013ApJ...766..115K} estimate a lower limit for the radial accretion rate on Serpens South (with
a line mass of $\sim 60$~M$_\odot$~pc$^{-1}$) of $130$~M$_\odot$~Myr$^{-1}$, corresponding to a line mass accretion rate of
$390$~M$_\odot$~pc$^{-1}$~Myr$^{-1}$ at a filament length of $0.33$~pc.

Yet, some of the model accretion rates are substantial (e.g., model "trb p$_0$", and most of the magnetized models), to the point of
being unrealistically high. For realistic background densities, the volume (strictly speaking, the cross section area) from which the
filament would have to draw mass would extend out to a "radius of influence", 
\begin{equation}
  R_i = \sqrt{\frac{m}{\pi\rho}},\label{e:rinf}
\end{equation}
of several parsecs for $\dot{m}>100$~M$_\odot$~pc$^{-1}$~Myr$^{-1}$. Given the spatial
scales and the background column densities seen in observations, this seems unrealistic.
These high accretion rates are mostly a consequence of the filament's expansion. 
The filaments expand most for models with constant external pressures, i.e. for cases where the infalling gas does not exert a corresponding
ram pressure. In other words, the unrealistic results for these cases are a consequence of an inconsistency in the model setup -- if we assume
infall at multiples of the sound speed and non-negliglible densities, this mass flow should be dynamically important and "squeeze" the filament,
reducing the filament radius, and, hence, lowering the mass accretion rate. 

\subsubsection{Steady-State Accretion}\label{sss:steadystate}
The models rest on the assumption of steady-state, free-fall mass accretion (eq.~\ref{e:vrad}), where the velocity profile $v_R$ evolves 
with the line mass, and thus with time. Any information about a change in the line mass travels
over the "radius of influence" (eq.~\ref{e:rinf}) on time scales much shorter than the flow time scales. While inconsistent
with the free-fall assumption (since the flow time scale is the same as the crossing time scale), it serves as a necessary simplification
short of solving the fully time-dependent, hydrodynamical problem. This inconsistency will lead to an over-estimate of the mass accretion,
and thus render evolutionary timescales as lower limits. 

Starting with a fully-developed accretion profile at $t=0$, instead of gas at rest, may seem unrealistic, yet it is motivated by the need to
actually form a seed filament by e.g. shock compression \citep{2000ApJ...535..887K,2001ApJ...553..227P}, 
possibly including thermal and shear effects \citep{2013A&A...556A.153H}. 
Such a scenario of filament formation would require inflows, most likely at magnitudes
higher than the assumed initial inflow velocities, which range around the value of the isothermal sound speed (Fig.~\ref{f:plotsingle_all}).

Given that the ambient medium is pressurized, free-fall can only be assumed for $v(R_f)\gg c_s$. This stage is reached for most models 
at $f>0.3$. Though this assumption is questionable for earlier stages, the choice of $f$ (as discussed in H13) just sets the initial mass
of the filament, and thus depends somewhat on the filament formation scenario. In other words, assuming an instantaneous
formation of a free-fall profile (see preceding discussion) allows us to neglect the flow history, and thus choose any value of $0<f<1$ as
starting point.

Finally, given the initial conditions of a low-mass filament as ``seed'', and with mass in the ambient medium than in the filament, 
one might argue whether the flows onto the filament would be more appropriately described as ``global collapse'' rather than ``accretion''. 
I use the term ``accretion'' here to emphasize a connection to observed filaments, which are usually seen at an evolved stage. 

\subsection{Filament Evolution and FWHM($N_c$)-Correlations}\label{ss:filevol}
In their Herschel study of dust filaments in Aquila, Polaris and IC5145, \citet{2011A&A...529L...6A} point out
that the filament width does not depend on the central column density $N_c$, even for the thermally gravitationally unstable
filaments in their sample (i.e. $f>1$ for $T=10$K. These will be referred to as "nominally unstable"). They argue  
that a turbulent filament formation mechanism as discussed by \citet{2001ApJ...553..227P} may explain the similar widths for
nominally {\em stable} filaments (i.e. $f<1$), and that for nominally {\em unstable} ($f>1$)
filaments it could be a consequence of continuing accretion, assuming a virialized filament. 
%In their study of externally pressurized, isothermal filaments, FM12a noticed that 
%that such nominally unstable filaments could be achieved for small inclination angles $\cos i<1$, though they
%point out that extremely high column densities would be needed. In their follow-up paper, FM12b show that a
%sample of four low-mass filaments (two of which also were studied by \citet{2011A&A...529L...6A}) show 
%profiles and parameters consistent with the model of externally pressurized filaments developed in FM12a.

H13 explored the role of turbulence on the filament accretion and subsequent evolution, finding that
accretion-driven turbulence can in principle lead to a decorrelation of $N_c$ and FWHM, as speculated by \citet{2011A&A...529L...6A}.
To test whether the models discussed in this study are consistent with the previous results (and with observations), 
$500$ filament accretion models were run for each case considered (see Table~\ref{t:sampleparam}).

\begin{table*}
\begin{center}
\caption[Filament Parameters]{Parameter ranges are indicated by $\dots$. For non-constant values, the relevant
equation numbers are given.\label{t:sampleparam}}
\begin{tabular}{lcccccc}
\hline
%\tablenotemark{a}
model       & $T_0$      & $\sigma$               & $\epsilon$                   & $\beta_0$           & $s$ & $p_{ext}$           \\
            & [K]        &                        &                              &                     &     & [$10^4$~K~cm$^{-3}]$\\
\hline
iso p$_0$   & $5\dots15$ & $c_s$                  & --                           & --                  & --  & $2$                 \\
iso p$_v$   & $5\dots15$ & $c_s$                  & --                           & --                  & --  & eq.~\ref{e:pext}    \\
trb p$_0$   & $10$       & eq.~\ref{e:sigmaturb}  & $5\times10^{-3}\dots 10^{-1}$& --                  & --  & $2$                 \\
trb p$_v$   & $10$       & eq.~\ref{e:sigmaturb}  & $5\times10^{-3}\dots 10^{-1}$& --                  & --  & eq.~\ref{e:pext}    \\
m05 p$_0$   & $10$       & eq.~\ref{e:magsound}   & --                           &$10^{-1}\dots10^2$   &$0.5$& $2$                 \\
m10 p$_0$   & $10$       & eq.~\ref{e:magsound}   & --                           &$10^{-1}\dots10^2$   &$1.0$& $2$                 \\
m05 p$_v$   & $10$       & eq.~\ref{e:magsound}   & --                           &$10^{-1}\dots10^2$   &$0.5$& eq.~\ref{e:pext}    \\
m10 p$_v$   & $10$       & eq.~\ref{e:magsound}   & --                           &$10^{-1}\dots10^2$   &$1.0$& eq.~\ref{e:pext}    \\
\hline
\end{tabular}
\end{center}
%\tablenotetext{a}{ Assume the kinematic ``near'' distances from \citet{ragan_msxsurv}.}
%\tablenotetext{b}{ of the combined data set.}
\end{table*}

\begin{figure*}
  \includegraphics[width=\textwidth]{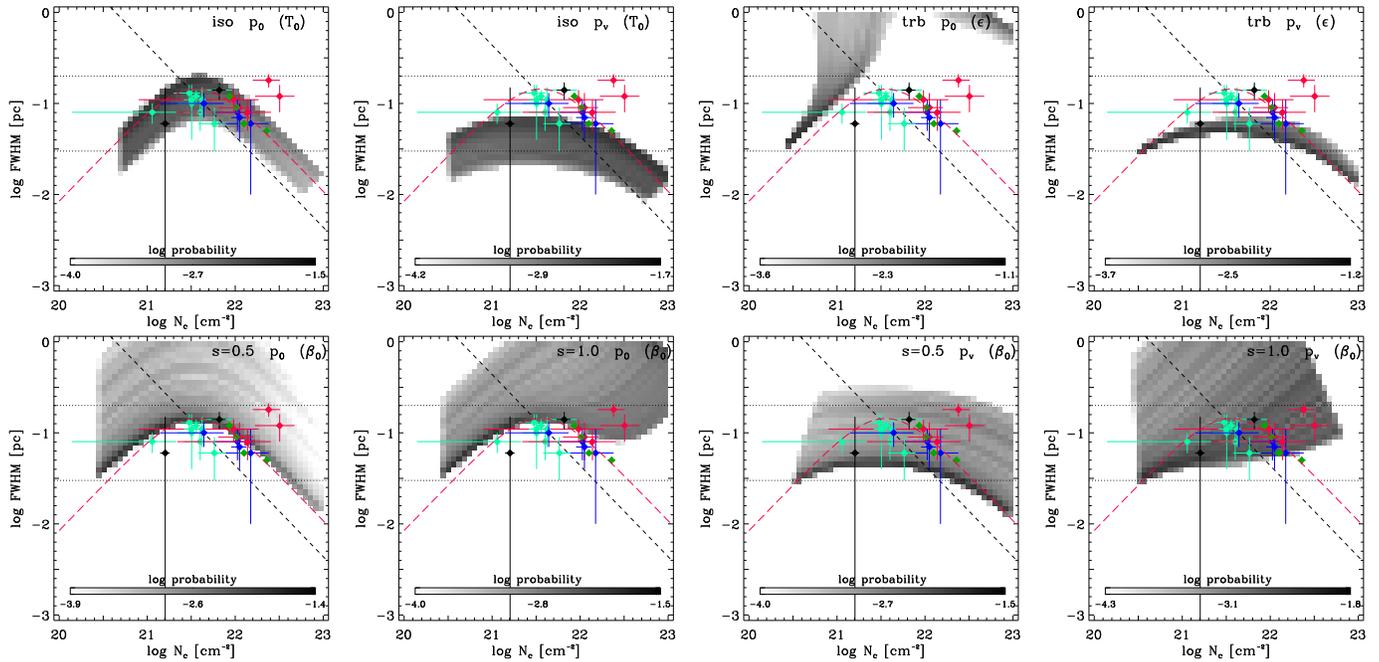}
  \caption{\label{f:arzoumanian}Probability density map of FWHM against atomic column density for all cases considered in \S\ref{s:accretion}. 
           The relevant physical parameters are indicated in each panel, including the parameter that is varied for each set of models 
           (in parentheses), and the sampling ranges are given in Table~\ref{t:sampleparam}.
           The plots can be directly compared to Figure~7 of
           \citet{2011A&A...529L...6A}, or Figure~7 of FM12b. 
           The range of observed FWsHM is given by the dotted lines, and the dashed line indicate the Jeans length,
           Observed filament values are plotted as diamonds, with colors indicating whether the filament contains YSOs (red), pre-stellar
           cores (blue), cores (light green), or nothing (black). Dark green diamonds indicate the values derived by FM12b.}
\end{figure*}

Figure~\ref{f:arzoumanian} summarizes the test results. Each panel shows the probability to find a filament at a given
position in $(\mbox{FWHM},N_c)$ space during its evolution. Evolutionary tracks follow the overall envelope shapes, and
in some cases, they are visible because of the finite sampling. The eight panels correspond to the eight cases
summarized in Table~\ref{t:sampleparam}. I overplotted the FWHM($N_c$) values for selected filaments in IC~5146, 
drawn from \citet[][their Table~1]{2011A&A...529L...6A}. Colors indicate whether the filament
contains YSOs (red), pre-stellar cores (blue), cores (green), or nothing (black).
Overall, only a few of the eight cases allow a wide enough range in FWsHM consistent with the
observed decorrelation between FWHM and $N_c$. I show the {\em atomic} central column density, consistent with FM12b, and 
in difference to \citet{2011A&A...529L...6A}. Latter authors use the {\em molecular} column density, though their mean
atomic weight seems to be inconsistent with that choice (see footnote in FM12b).

The isothermal case (``iso p$_0$'') reproduces Figure~10 of FM12a for a temperature of $T=10$~K (red dashed line). 
This is expected, yet
it serves here as a consistency check.
The tracks follow the instability line for large $N_c$, but, by construction, only
stable filaments are allowed. 
FM12a discuss the possibility of projection effects (or viewing angles) to shift the evolutionary tracks to higher
column densities, and thus break the strong FWHM$(N_c)$-correlation, although they demonstrate that in this case, the high-column density filaments
all would be seen at rather extreme inclination angles.
Other possibilities would involve a non-uniform column density along the backbone of the filament (i.e. cores), pushing the 
average along the filament to higher values, or filaments embedded in a background medium of non-zero column density (Sec.~\ref{s:sheetfilaments}).
Column densities and FWsHM derived by FM12b for a sample of four filaments are consistent with model "iso p$_0$" (dark green symbols
in Fig.~\ref{f:arzoumanian}). Note that two of their filaments have also been analyzed by \citet{2011A&A...529L...6A}, and that they arrive
at different parameters. This may be just due to the different analysis methods, as discussed by FM12b.

If the external pressure contains a ram-pressure component (case ``iso p$_V$'') due to the varying accretion velocity (eq.~\ref{e:pext}), the 
filament gets ``squeezed'', and all characteristic scales shrink. The FWsHM now mostly lie below the observed range. The overall 
distribution is flatter than for the isothermal case with constant pressure, suggesting that varying pressures can lead to a decorrelation 
between $N_c$ and the FWHM. Yet, since free-fall accretion has been assumed, the external ram pressure will be a generous upper 
limit, suggesting that for lower and more realistic ram pressure estimates, {\em the distribution of filament trajectories could move towards 
higher FWsHM}. This would also reduce the effect of the ram pressure over the external pressure, thus increasing the curvature of 
the trajectories again, and thus converging to the constant pressure case. Only if the external constant pressure component
were reduced substantially (by a factor of $10$), the trajectories could be shifted into the observed FWHM range.

The turbulent case with constant external pressure (``trb p$_0$'') does not require much discussion: 
the unconstrained filament is rapidly expanding, even seemingly avoiding
the observed parameter ranges (see also \S\ref{sss:massreservoir}). Adding a ram pressure component again pressurizes the filament (``trb p$_V$''). 
The FWHM distribution is much narrower than for the isothermal case, because of the lower power with which the filament 
properties depend on the turbulent driving efficiency (eq.~\ref{e:sigmaturb}). The distribution extends into the nominally (isothermally) unstable
regime and has flattened, yet, the values lie again below most of the observed parameters.

Introducing a magnetic field assuming mass and flux conservation (``$s=0.5$ p$_0$'') results in a trajectory distribution similar to that of the isothermal
case, since the magnetic field for $s=0.5$ just contributes a constant addition to the sound speed (eq.~\ref{e:magsound}). A decorrelation of $N_c$ and
FWHM only would be expected if filaments at low and high central column densities are less magnetized than those at intermediate $N_c$ -- a not entirely
convincing scenario. The situation improves drastically when adding the ram pressure to the external pressure (``$s=0.5$ p$_V$''). The trajectories are
now fairly flat, with the FWHM depending only weakly on $N_c$, and they are approximately within the observed range. 

A poloidal or toroidal field (``$s=1.0$ p$_0$'' and ``$s=1.0$ p$_V$'') also leads to a decorrelation between FWHM and $N_c$, yet, due to the
magnetization now increasing with the filament density, the filament scales increase well above the observed range. One could speculate whether such broad
filaments could be missed in the observational analysis, since they essentially would appear as background. Yet note that these cases are equivalent to
an effective equation of state with $\gamma=2$, since $B\propto n$.    

From the above discussion, we conclude the following:
\paragraph{(a)} The isothermal case as discussed by FM12a is unlikely to show a decorrelation of FWHM and $N_c$ to the extent observed
by \citet{2011A&A...529L...6A}. 
Especially when approaching the instability line (dashed diagonals in Figure~\ref{f:arzoumanian}), the correlation becomes inevitable 
just by construction. The column densities and FWsHM derived by FM12b (their Fig.~7) for low-mass filaments 
are consistent with the isothermal case (dark green symbols, dashed red line).
\paragraph{(b)} The ram-pressurized, isothermal case, and the turbulent cases 
(while still assuming -- in contrast to H13 -- $p=4$) cannot break the correlation, 
unless unrealistically low external constant pressure components are assumed.
\paragraph{(c)} Magnetic fields scaling linearly with density (i.e. an effective $\gamma=2$) lead to a decorrelation, but also to filament
widths substantially larger than observed. 
\paragraph{(d)} The best agreement is found for magnetic fields scaling weakly with density, especially if the external pressure contains a 
ram pressure component. Since $s=1/2$ is equivalent with mass and flux conservation, this result is consistent with the filaments embedded in
three-dimensional extended structures of more diffuse, magnetized material.
Reasonably, one could assume this to be the most realistic case also. \citet{2008ApJ...680..420H} found striations in the molecular
gas to be aligned with magnetic field vectors inferred from polarimetry in Taurus, consistent with kinematic evidence. 
The same pattern has been observed in much greater detail in dust column density maps \citep{2013A&A...550A..38P}. Such structures are 
consistent with a radial field component (in the plane of sky) around an accreting filament. From
field geometry considerations, the field is probably not perfectly radial for all azimuthal directions around the filament.
Yet, if the filament were embedded in a flattend cloud, the observed field directions would be also restricted to the plane of the cloud. This is
the situation envisaged for the case $s=0.5$ discussed above.

\section{Filaments within Sheets}\label{s:sheetfilaments}

\begin{figure}
  \includegraphics[width=\columnwidth]{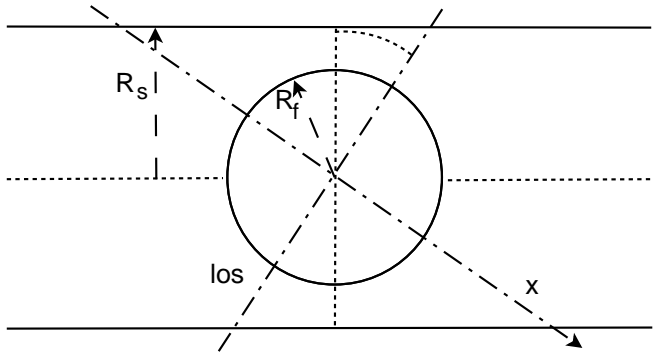}
  \caption{\label{f:sketch}Sketch of the geometry of a filament of radius $R_f$ embedded in a sheet of thickness $2R_s$ (case $R_f<R_s$ shown). 
           The projection angle $\cos\beta$ is measured between the line of sight and the normal onto the filament and plane, as indicated.}
\end{figure}

Given the observed column density and magnetic field structures around molecular filaments, it is not unreasonable to assume that the 
filaments are embedded within structures of the next higher dimensions, i.e. within flattened clouds, or sheets. \citet{2011ApJ...740...88P} discuss
a sequence of gravitational collapse, from higher to lower dimensions, consistent with filaments embedded in sheets. 
Flattened structures result also naturally from clouds forming in large-scale flows. In the following, I develop a simple model of a filament embedded in a
sheet, with the goal to explain the flattening of observed filament profiles in comparison with the hydrostatic cylinder. 
This might be considered a logical extension of the discussion by FM12a: not only is it realistic to assume that the filament is embedded
within a medium of non-zero pressure, but also of non-zero density. The latter effect will be discussed here, with the goal
to develop an observational diagnostic identifying the evolutionary stage of the filament (parameterized by the criticality parameter $f$, 
eq.~\ref{e:fcrit}), and the projection angle between the line-of-sight and the embedding sheet (see Fig.~\ref{f:sketch}).

\subsubsection{Flattening of Profiles}\label{sss:flatprofiles}

\citet{2011A&A...529L...6A} conclude 
that their identified filaments have radial column density profiles generally flatter than 
those expected for an isothermal, infinite cylinder \cite[see also][]{2013A&A...550A..38P}.
FM12a explain the flatness of the profiles with external pressurization: only for zero external
pressure, the filament will show a density profile $\propto R^{-4}$ (see their Fig.~3). 
Here, I explore the effect of a non-vanishing background column density on the derived filament properties, specifically on
the steepness of the profile, parameterized by the power law index $p$ (see eq.~\ref{e:rhoRofp}). A vanishing background 
density would be fully appropriate only if the temperature outside the filament is high enough to render the ambient density (and thus column
density) below the observational sensitivity. 

For the following analysis, I consider an externally pressurized filament of radius $R_f$ embedded in a sheet of thickness $2R_s$. 
Thus, if $R_s=R_f$, the filament fits snugly into the sheet\footnote{I neglect any scale heights here and assume a top-hat for the
filament profile. Adding a scale height would introduce more parameters, but not qualitatively change the results}. 
We are interested in the total column density profile $N_{tot}(x)$ through
the filament and the ambient sheet, depending on the viewing angle $\cos\beta$, with $\cos\beta = 1$ indicating a line-of-sight perpendicular
to the sheet, and $\cos\beta=0$ parallel to the plane defined by the sheet, and perpendicular to the filament (see Fig.~\ref{f:sketch}). 
For $\cos\beta=1$, $N_{tot}(0) = N_c$. The radius $x$ measures the distance in the plane-of-sky, normalized to the filament radius $R_f$. 
The column density of the filament $N_f(x)$ is given by eq.~15 of FM12a, where the constant external pressure $p_{ext,0}$ is replaced by
equation~\ref{e:pext}. The column density of the ambient gas is  
\begin{eqnarray}
  N_s(x)&=&2 n_{ext}\left(\frac{R_s}{\cos\beta}-R_f(1-\bar{x}^2)^{1/2}\right),\mbox{  with}
  \label{e:ncolsheet}\\
  \bar{x}&\equiv& \min(1,x).
\end{eqnarray}
Thus, we can express the total column density $N_{tot}=N_f(x)+N_s(x)$ for sheets of varying thickness, and viewed under arbitrary angles.
Obviously, eq.~\ref{e:ncolsheet} only holds if the embedding sheet extends at least a distance 
\begin{equation}
  D_{min}=\frac{R_s/R_f}{\tan(\pi/2-\beta)}
\end{equation}
from the filament center.

The above equations allow us to generate a set of filament models in dependence of $f$ and $\cos\beta$. For each pair $(f,\cos\beta)$, a column density
profile results that is fit with the profile given  by Arzoumanian et al. (2011, their eq.~1), resulting in a flatness parameter $p$,
a core radius $R_0$ (their $R_{flat}$), and the ratio of the column density over the background column
\begin{equation}
  Q_N\equiv N_{tot}(0)/N_s(R_f).\label{e:QN}
\end{equation}
The fit parameters have errors less than $3$\% for $f<0.99$ and $\cos\beta>10^{-2}$, i.e. they
are useful for reasonable ranges of $f$ and $\cos\beta$.

\begin{figure*}
  \includegraphics[width=\textwidth]{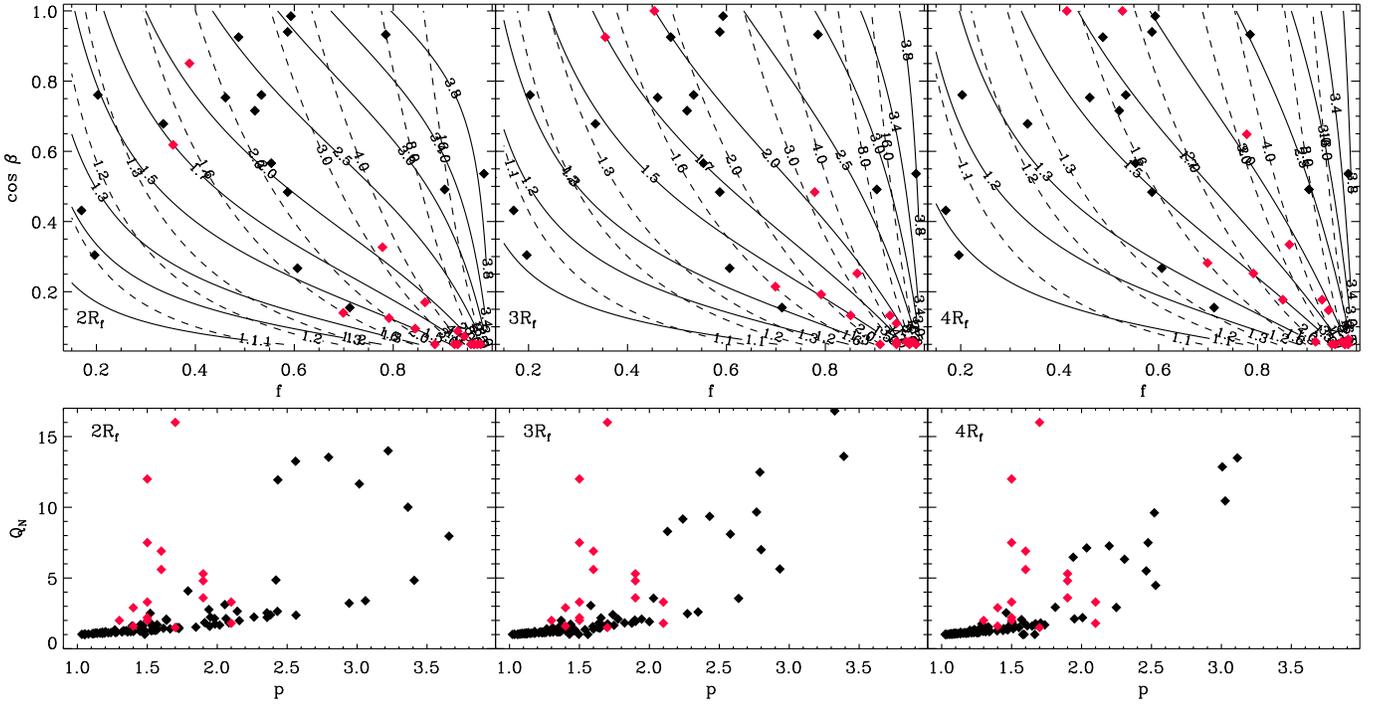}
  \caption{\label{f:contpNr}{\em Top:} Contour lines of the profile exponent $p$ (solid lines) and the central to background 
           column density $Q_N$ (eq.~\ref{e:QN}, dashed lines) in terms of the criticality parameter $f$ and the viewing angle $\cos\beta$ (see text).
           If $p$ and $N_c/N_{ext}$ can be determined from observations, the three diagrams provide estimates for the evolutionary stage of
           the filament in terms of $f$, and the viewing angle $\cos\beta$, for a filament as modeled above embedded in a sheet of thickness $2,3$, and $4R_f$
           (left to right, indicated in bottom left of each panel). All three diagrams use the same set of contour levels. For $p$, these are
	   given by $p=[1.1,1.2,1.3,1.7,2.0,2.5,3.0,3.4,3.8]$, and $N_c/N_{ext}=[1.1,1.2,1.3,1.6,2.0,3.0,4.0,8.0,16.0]$. For increasing embedding
           sheet thickness, the contours shift to higher $f$. {\em Bottom:} Distribution of observed filaments (red symbols; 
           taken from Table~1 of Arzoumanian et al. 2011)}
\end{figure*}

If the exponent $p$ and the column density ratio $Q_N$ are known from observations, then
the projection angle $\cos\beta$ {\em and} the criticality parameter $f$ of the pressurized filament can be estimated by minimizing 
\begin{equation}
  \Delta^2_{f,\beta}\equiv \left( \left(\frac{Q_N(obs)-Q_N(f,\cos\beta)}{Q_N(f,\cos\beta)}\right)^2
                                 +\left(\frac{p(obs)-p(f,\cos\beta)}{p(f,\cos\beta)}\right)^2\right)
  \label{e:minimize}
\end{equation}
over a map of $(f,\cos\beta)$. The subscript $obs$ indicates the observed values.
The top row of Figure~\ref{f:contpNr} indicates how reliable such estimates might be. It shows contour lines of $p$ (solid lines) and $Q_N$ (dashed line), in the
$(f,\cos\beta)$-plane, for three values of $R_s/R_f=1,1.5,2$. The contours are labeled with their respective values. The central column density depends on 
$f$ through eq.~15 of FM12a, while the total column through filament and sheet depends on the angle $\cos\beta$ via eq.~\ref{e:ncolsheet}.
The black symbols plotted over the contour lines are used as a check how reliably eq.~\ref{e:minimize} can retrieve the parameters. A $100$ randomly chosen
points, uniformly distributed in $\cos\beta$ and $f$, were used to calculate the corresponding $Q_N$ and $p$ values. For the physically interesting
ranges of $f$ and $\cos\beta$ as mentioned above, the values are recovered with less than $3$\% error, not surprisingly consistent with the fit accuracies.

The method works best for thin sheets -- since the sheet column density does not depend on $f$, angles between contours of $p$ and $Q_N$ are larger for
larger $f$, and thus degeneracies between $f$ and $\cos\beta$ are less likely to occur. With increasing $R_s/R_f$, the curves start to align, and thus 
the estimates of $\cos\beta$ and $p$ will be correspondingly less certain.

Taking the column density values and $p$ values from Table~1 of Arzoumanian et al. (2011) and applying equation~\ref{e:minimize} 
generates the red symbols in Figure~\ref{f:contpNr}. We note the following issues: (a) With increasing $R_s$ (center and right panel), 
the observed points move towards $\cos\beta=1$. This is just because for increasing sheet thickness, the contrast $Q_N$ drops. While 
we cannot determine $R_s$ independently, one could argue from the distribution in $\cos\beta$ that larger $R_s$ values may be 
unrealistic, since they would entail $\cos\beta=1$ for most filaments. (b) Most of the observed points cluster at $0.9<f<1$ and 
small $\cos\beta$. Taken at face value, this would mean that for most of the filaments, the line-of-sight is nearly parallel to 
the embedding sheet. This may (or may not) be an unsatifactory conclusion, and could be tested with kinematic data. Yet, it should 
be pointed out that all those points correspond to fairly high $Q_N$ and low $p$, a region of the parameter space where the curves 
of constant $N$ and $p$ are nearly parallel. Also, the models used to generate the $(p,Q_N)$-curves assume that the actual density 
profile corresponds to an isothermal cylinder, with $p=4$ in vacuum. 

The clustering\footnote{Note that the clustering is a consequence of the minimization (eq.~\ref{e:minimize}). Strictly speaking,
no viable solutions are being found in that regime of parameters.} 
of the observed value in a nearly inaccessible parameter region suggests to check the underlying distribution of 
$(p,Q_N)$-values for the observations (red symbols in Figure~\ref{f:contpNr}, bottom) and test models (black symbols). The difference 
between both distributions is obvious: randomly choosing values in $\cos\beta$ and $f$ does not lead to a full coverage of 
$(p,Q_N)$-space, as already evident from the contour distribution in Figure~\ref{f:contpNr}. Yet, these are the values that are 
``allowed'' in this simple model, assuming a hydrostatic cylinder embedded in a sheet. While the modeled values can reach high $Q_N$ only for
large $p$, the observed values are clustered at lower $p$, with similarly high $Q_N$ values. Only the low-$Q_N$ regime shows overlap. 
Increasing $R_s$ closes the gap between the two distributions somewhat, yet, it also reduces the $Q_N$ values modeled. If the 
modeled $Q_N$ are to be kept at similar levels, larger $f$ are required. 

Summarizing, the observed regime of high $Q_N$ and low $p$ {\em is not accessible by an isothermal, externally pressurized 
cylinder embedded in a sheet}. Low $p$ would entail a small $\cos\beta$ in the model (and a small $Q_N$), since the profiles 
flatten with decreasing $\cos\beta$ and $f$. To drive up the contrast to observed 
values, a correspondingly larger value of $f$ would have to be chosen. 
The reason for the failure lies in the fact that the intrinsic (3D) profile still has $p=4$, and the flattening is solely due to a 
lower $f$ (and thus larger $R_0$) and/or a lower $\cos\beta$ or higher background column density. Thus, low $p$ is interpreted
as a low $f$ and hence a low $Q_N$, whereas a low $p$ and {\em high} $Q_N$ would require extremely small $\cos\beta$ or high backgrounds
to ``hide'' the over-density.

While the described method to estimate the evolutionary stage of a filament and its 
environment fails for the isothermal cylinder, the technique might prove useful for more generalized cylinder models, specifically for models with flatter
intrinsic profiles.

\subsubsection{Temperature Profiles}
Here, we briefly speculate on the effect of different temperatures in the filament and the embedding sheet on the overall temperature profile 
of an observed filament in dependence of the viewing angle $\cos\beta$. Two constant temperatures for the sheet, $T_s=15$~K, and for
the filament, $T_f=10$~K, are assumed. The temperature profile is then integrated along the line-of-sight set by $\cos\beta$ as
\begin{equation}
  T(x) = \frac{N_f(x)T_f+N_s(x)T_s}{N_{tot}(x)},
\end{equation} 
with the projected distance to the filament center $x$, measured in units of $R_f$. 

\begin{figure*}
  \includegraphics[width=\textwidth]{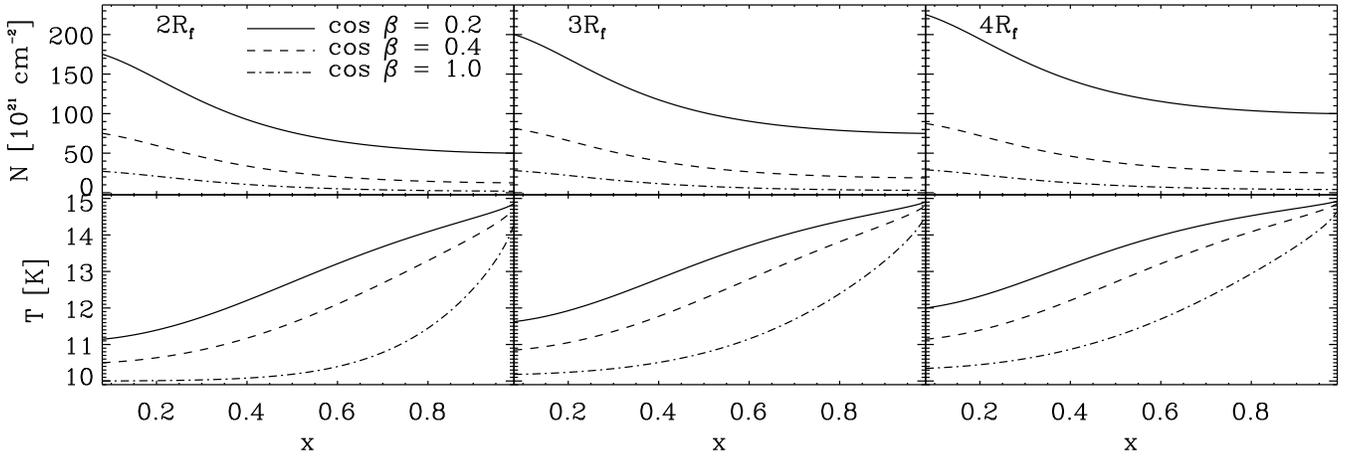}  
  \caption{\label{f:tempprof}Column density (top) and temperature (bottom) profiles of a pressurized, hydrostatic, isothermal cylinder
           at $f=0.8$, embedded in a sheet of uniform density $n_{ext}$ and thickness of $2,3$, and $4R_f$ (left to right, as indicated in panels).
           Temperatures in the sheet and filament are assumed to be constant, at $T_s=15$~K and $T_f=10$~K, respectively. Assuming density-weighted
           temperatures, already this simple geometry can lead to temperature profiles consistent with observations.} 
\end{figure*}

The results for the three values of $R_s=1.0,1.5,2.0$ are summarized in Figure~\ref{f:tempprof}. Already this simple temperature distribution
and geometry can generate temperature profiles consistent with observations. Note that this is not intended to suggest that observed filaments 
are isothermal. The only point being made here is to show that projection effects can have
a substantial effect on the derived temperature, of the same order as intrinsic temperature variations.

%%%%%%%%%%%%%%%%%%%%%%%%%%%%%%%%%%%%%%%%%%%%%%%%%%%%%%%%%%%%%%
\section{Conclusions}
In an extension of \citet{2012A&A...542A..77F} and \citet{2013ApJ...769..115H}, the evolution of a
pressurized, hydrostatic cylinder accreting gas at free-fall velocities is studied. FM12a discussed 
the problem of an externally pressurized, isothermal cylinder, and how such a model would explain Herschel
observations of molecular cloud filaments and their stability (FM12b). H13 explored the effect of accretion on a 
filament described by a central density, a core radius, and a power-law profile with exponent $p$, 
following the parameterization by \citet{2011A&A...529L...6A}. 
%Both studies could explain the observed 
%decorrelation between the filament FWHM and the central column density $N_c$, in one case through projection 
%effects (FM12a), and in the other through accretion-driven turbulence (H13). 

Here, the goal is to merge the two models. The main difference to FM12a is that the external pressure is 
allowed to vary due to the ram pressure exerted by the infalling gas (eq.~\ref{e:pext}). This additional 
pressurization leads to a compression of the filament. Unlike H13 (and following FM12a), a physical model 
for the filament is assumed, namely a hydrostatic isothermal cylinder, thus motivating the choice of $p$. 

The accretion models (Fig.~\ref{f:plotsingle_all}) reproduce the results by FM12a for the isothermal case. 
Ram-pressurized, hydrodynamical accretion models lead to filament FWsHM an order of magnitude smaller than 
observed (Fig.~\ref{f:arzoumanian}), suggesting that (a) the ram pressure is being over-estimated, due to the 
strong free-fall assumption (see discussion by H13) or (b), that -- since H13 could explain the decorrelation 
between FWHM and $N_c$ for filaments flatter than isothermal -- the isothermal cylinder is not a fully 
adequate description for currently observed filaments. Yet, FM12b demonstrate that for their sample of
low-filaments (two of which are re-analyzes of filaments discussed by \citet{2011A&A...529L...6A}), the isothermal
cylinder provides a good description for the observed $N(FWHM)$ values. These differences in the 
interpretation of observational data may well still be due to the specific analysis techniques and selection
of filament (sections) used by each of these authors\footnote{The referee points out that the observational 
results of \citet{2011A&A...529L...6A} might be biased by
the method the data had been analyzed. The profile in their model was an average along filamentary structures
where the centre of the filament was assumed to be given by the maximum along the different cuts. If the massive
filaments contained embedded structures, this would result in a well-defined inner profile surrounded by a less defined
outer envelope. The profile shown in  Fig.~4a of \citet{2011A&A...529L...6A} belonged to the massive structure in
IC 5146, called also the 'Northern Streamer'. The image in their paper showed a rather complex structure not expected
for isothermal filaments. Yet, it also is worth mentioning that FM12b restrict their analysis to small filament sections
that show little background and little confusion from nearby sources, and that thus their sample may be biased also.}

While magnetized accretion models other than the simplest, magnetosonic case can reproduce a decorrelation, 
cases with strong magnetic scaling $B\propto n$ result in a FWHM distribution substantially broader than 
observed. Only the ram-pressurized weak scaling case shows some promise to reproduce observed FWHM$(N_c)$ 
values adequately. This seems a reasonable conclusion in face of observed column density striations aligned 
with magnetic field vectors around Taurus molecular filaments \citep{2008ApJ...680..420H,2013A&A...550A..38P}, 
if a filament-in-sheet geometry is assumed. 

This filament-in-sheet geometry is explored to some extent in \S\ref{s:sheetfilaments}. Extending FM12a's 
reasoning and assuming not only a non-zero ambient pressure, but also a non-zero ambient density, suggests 
that in the simplest case, a pressurized filament would sit in a sheet of some thickness equal to or larger 
than the filament diameter (Fig.~\ref{f:sketch}). Such a configuration would lead to a natural flattening of 
the profile, in addition to a flattening introduced by reducing the evolution parameter $f$. Given observed 
profile parameters $p$ and column density contrasts $Q_N$ (eq.~\ref{e:QN}) between central and ambient column 
density, the viewing angle $\cos\beta$ and the evolution parameter $f$ can be estimated (Fig.~\ref{f:contpNr}). 
Testing this method with the data from  \citet{2011A&A...529L...6A} demonstrates that the observed 
parameter regime of high $Q_N$ and low $p$ is not accessible by a simple filament-in-sheet geometry, for any 
kind of ratio $R_s/R_f$, {\em if the cylinder is isothermal}.
%This suggests that the profiles observed by \citet{2011A&A...529L...6A} may be 
%intrinsically flatter than isothermal, beyond projection and evolution effects. Yet, it should be noted
%that FM12b do find agreement of their low-mass filaments with an isothermal cylinder, especially for two of the filaments also analyzed
%in Arzoumanian's et al. sample. 

\acknowledgements 
The anonymous referee's report was extremely thorough, insightful, and helpful. Thank you very much indeed.
This study originated from a comment the anonymous referee made on the manuscript of Heitsch (2013). It is 
also a spin-off from discussions and a presentation given at the Early Phases of Star Formation 2012 meeting 
in Ringberg, organized by J.~Steinacker and A.~Bacmann. 
I gratefully acknowledge support by NSF grants AST-0807305 and AST-1109085. 

\clearpage

\newpage

\appendix

\section{Summary of Fischera \& Martin (2012a)}\label{a:fischera}
The relevant expressions of \citet[][see their section 2]{2012A&A...542A..77F} are briefly summarized.
The density profile of an isothermal, hydrostatic, infinite cylinder is given by
\begin{equation}
  \rho(R) = \rho_c\left(1+\left(\frac{R}{R_0}\right)^2\right)^{-2}
  \label{e:rhoR}
\end{equation}
\citep{1964ApJ...140.1056O}, with the core radius $R_0$ given by equation~\ref{e:R0}.
The corresponding line mass in dependence of the filament radius is
\begin{equation}
  f(R_f)=\frac{m}{m_{cr}} = 1-\frac{1}{1+R_f^2/R_0^2},
  \label{e:linemass}
\end{equation}
and thus, with $p(R_f)=\rho(R_f)c_s^2=p_{ext}$,
the central density in terms of the external pressure is given by
\begin{equation}
  \rho_c=\frac{p_{ext}}{c_s^2} (1+R_f^2/R_0^2) = \frac{p_{ext}}{c_s^2}\frac{1}{(1-f)^2}.
  \label{e:rhoc}
\end{equation}
Using equations~\ref{e:R0} and \ref{e:linemass} results in
the filament radius
\begin{equation}
  R_f = c_s^2\left(\frac{2f(1-f)}{\pi G p_{ext}}\right)^{1/2}.
\end{equation}

As \citet{2012A&A...542A..77F} show in their Figure~3, the overpressure $p_c/p_{ext}$
will determine the steepness of the resulting density profile: for large over-pressures
(or $R_f/R_0\gg 1$), the profiles approach the vacuum solution $\propto R^{-4}$.
Thus, the overpressure will determine the "flatness" of the cylinder
profile.

\section{Pressurized Filaments: Expressions for Root Finders}\label{a:rootfinder}

I summarize the expressions and methods to derive the filament accretion rate, assuming an externally pressurized,
hydrostatic filament.
To integrate the line mass density (eq.~\ref{e:dmdt}), the filament radius $R_f$ (eq.~\ref{e:radmain}), the criticality
parameter $f$ (eq.~\ref{e:fcrit}), and the sound speed $\sigma$ are needed in the RHS of the ODE. For most of the
cases, $R_f$ depends on itself in a non-trivial way, through the ram pressure or through the sound speed. The approaches
to resolve these dependencies are discussed in the following. The full expressions and their implementation can be found 
at {\tt www.physics.unc.edu/$\sim$fheitsch/codesdata.php}

\subsection{Constant $\sigma$, varying external pressure}\label{aa:iso1}
This is the case "iso p$_{\mbox{v}}$" in Figure~\ref{f:plotsingle_all}. The external pressure in eq.~\ref{e:radmain} is replaced
by the combination of ram pressure and constant thermal pressure, eq.~\ref{e:pram}, resulting in
\begin{equation}
  1-\f{\sigma^2}{R_f} \left(\f{2f(1-f)}{\pi G (p_{ext,0}+4 G m q \rho_{ext})}\right)^{1/2}\equiv 0,
  \label{e:iso1root}
\end{equation}
where 
\begin{equation}
  q\equiv \ln\frac{R_{ref}}{R_f}.
\end{equation}
Equation~\ref{e:iso1root} can be solved with a simple root finder.

\subsection{Accretion-driven turbulence, constant external pressure}\label{aa:iso2}
This is the case ``trb p$_0$'' in Figure~\ref{f:plotsingle_all}. The soundspeed $\sigma=c_s$ is replaced by equation~\ref{e:sigmaturb}.
It turns out to be advantageous in terms of accuracy and convergence speed to solve the dependencies with a Newton-Raphson method
for the scaled filament radius $\bar{R}\equiv R_f/R_{pc}$, where $R_{pc}$ is a scaling number. Thus, we get the following expression:
\begin{eqnarray}
  y(R)&=&\frac{\sigma^2}{\bar{R}R_{pc}}\left(\frac{2f(1-f)}{\pi G p_{ext,0}}\right)^{1/2}-1\label{e:iso2rootfinder}\\
      &=&uvw-1\equiv 0\nonumber
\end{eqnarray}
with 
\begin{eqnarray}
  u&=&\frac{1}{\bar{R}R_{pc}}\label{e:iso2u}\\
  v&=&\sigma^2\label{e:iso2v}\\
  w&=&\left(\frac{2f(1-f)}{\pi G p_{ext,0}}\right)^{1/2}\label{e:iso2w}.
\end{eqnarray}
The Newton-Raphson root finder also requires the derivative with respect to $\bar{R}$,
\begin{equation}
  y'=u'\,v\,w+u\,v'\,w+u\,v\,w',
  \label{e:derivuvw}
\end{equation}
with the derivatives
\begin{eqnarray}
  u'&=&-\frac{u}{\bar{R}}\label{e:dudR}\\
  v'&=&2\sigma\sigma'=a_2\bar{R}^{1/3}\left(\frac{4}{3}q-1\right),\label{e:dvdr}\\
  w'&=&\frac{1-2f}{(2\pi G p_{ext,0} f(1-f))^{1/2}}\,f'\label{dwdR},\\
  f'&=&-\frac{a_2 m G \bar{R}^{1/3}}{2\sigma^4}\left(\frac{4}{3}q-1\right).
\end{eqnarray}
Here, $a_2$ is a constant of the value
\begin{equation}
  a_2\equiv4\left(4\pi \epsilon (G^3 m)^{1/2} \rho_{ext} R_{pc}^2\right)^{2/3}.
\end{equation}

\subsection{Accretion-driven turbulence, varying external pressure}\label{aa:iso3}
This is the case ``trb p$_V$'' in Figure~\ref{f:plotsingle_all}. The soundspeed $\sigma=c_s$ is replaced by equation~\ref{e:sigmaturb} again,
and the external pressure is given by equation~\ref{e:pext}. As in the previous case (\S\ref{aa:iso2}), we use a Newton-Raphson
method for the scaled filament radius $\bar{R}\equiv R_f/R_{pc}$. The expressions for $u$ and $v$ are identical to those in 
equations~\ref{e:iso2u} and \ref{e:iso2v}, since they do not depend on the external pressure. For $w$ we now get
\begin{equation}
  w=\left(\frac{2f(1-f)}{\pi G p_{ext}}\right)^{1/2},
\end{equation}
and
\begin{eqnarray}
  w'&=&\left(\frac{1}{2 \pi G}\right)^{1/2} 
             \frac{p_{ext}(1-2f)f'-f(1-f)p'_{ext}}{w p_{ext}^2},\label{e:wprime}\\
  p'_{ext}&=&-4 G m \frac{\rho_{ext}}{\bar{R}}. 
\end{eqnarray}

\subsection{Accretion of magnetic fields, constant external pressure}\label{aa:iso4}
The magnetic case with constant external pressure requires replacing the sound speed by equation~\ref{e:magsound}. The expressions
are simple enough that we can treat both $s=1/2$ and $s=1$ in one branch (this will change in the next step), thus we get the following
expression for the root function (it is numerically more advantageous to solve for $\sigma$ instead of $R_f$) and its derivative:
\begin{eqnarray}
  y(\sigma) = \frac{\sigma}{c_s}-u\\
  y' = \frac{1}{c_s}-\frac{a_4 m G (2s-1)w^{-4s+1}}{u^{1/2}\sigma^3},
\end{eqnarray}
with the auxilliary functions
\begin{eqnarray}
  u&=&\left(1+a_4 w^{-2(2 s-1)}\right)^{1/2}\\
  w&=&1-f = 1-\frac{m G}{2\sigma^2}.
\end{eqnarray}
The constant $a_4$ is given by
\begin{equation}
  a_4 = \frac{2}{\beta_0}\left(\frac{p_{ext,0}}{\rho_{c0}c_s^2}\right)^{2s-1}.
\end{equation}

\subsection{Accretion of magnetic fields, varying external pressure, $s=1/2$}\label{aa:iso5a}
%filament-26/27
In this case, the effective sound speed (eq.~\ref{e:magsound}) simplifies to $\sigma^2=c_s^2(1+2/\beta_0)$.
We use a Newton-Raphson method to find the root for the (scaled) radius equation,
\begin{equation}
  y(\bar{R}) = 1-\frac{a_5}{\bar{R}}\,(2b_5\,q+p_{ext,0}),
\end{equation}
with the constants
\begin{eqnarray}
  a_5&=& \frac{\sigma^2}{R_{pc}}\left(\frac{2f(1-f)}{\pi G}\right)^{1/2}\\
  b_5&=& 4 \pi G \rho_{ext}m.
\end{eqnarray}
The derivative with respect to $\bar{R}$ is given by
\begin{equation}
  y' = \frac{a_5}{\bar{R}^2}\frac{2(b_5q+p_{ext,0})+b_5}{2(b_5q+p_{ext,0})^{3/2}}.
\end{equation}

\subsection{Accretion of magnetic fields, varying external pressure, $s=1$}\label{aa:iso5b}
%filament 56,57
Replacing the central density in equation~\ref{e:magsound} by equation~\ref{e:rhoc}, the effective soundspeed is
now given by
\begin{equation}
  \sigma = c_s\left(1+\frac{2}{\beta_0}\frac{p_{ext}}{p_{ext,0}(1-f)^2}\right)^{1/2},
  \label{e:sigma6}
\end{equation}
where $p_{ext}$ is defined by equation~\ref{e:pext}. Equation~\ref{e:sigma6} results in a cubic for $\sigma^2$, with the solution
\begin{eqnarray}
  \sigma^2&=&\frac{2^{1/3}a_6^2}{3D^{1/3}}+\frac{D^{1/3}}{3 2^{1/3}}-\frac{2^{1/3}b_6}{D^{1/3}}+\frac{a_6}{3}\label{e:sigma2}\\
  D&=&2a_6^3+\sqrt{27}\sqrt{4a_6^3c_6-a_6^2b_6^2-18a_6b_6c_6+4b_6^3+27c_6^2}-9a_6b_6+27c_6\\
  a_6&=&\frac{mG}{2}+c_s^2\left(1+\frac{2}{\beta_0}\frac{p_{ext}}{p_{ext,0}}\right)\\
  b_6&=&mG(mG/4+c_s^2)\\
  c_6&=&(mGc_s)^2/4.
\end{eqnarray}
With $\sigma$ in hand, we can write the root function for the scaled radius,
\begin{eqnarray}
  y(\bar{R}) &=& uvw-1\\
  u&=&\frac{1}{\bar{R}R_{pc}}\\
  v&=&\sigma^2\\
  w&=&\left(\frac{2f(1-f)}{\pi G p_{ext}}\right)^{1/2}.
\end{eqnarray}
The derivative is formally given by equation~\ref{e:derivuvw}. If we label the four terms of $\sigma^2$ in equation~\ref{e:sigma2} 
as $S_1,S_2,S_3,S_4$, the derivative of $\sigma$ is 
\begin{eqnarray}
  \sigma'&=&\frac{1}{2\sigma}(S'_1+S'_2+S'_3+S'_4)\\
  S'_1&=&\frac{2^{1/3}a_6}{3D^{1/3}}\left(2a'_6-\frac{aD'}{3D}\right)\\
  S'_2&=&\frac{1}{9\cdot 2^{1/3}}D^{-2/3}D'\\
  S'_3&=&-\frac{2^{1/3}b_6}{3}D^{-4/3}D'\\
  S'_4&=&\frac{1}{3}a'_6\\
  D'&=&\left(6a_6^2+\frac{3}{2}\frac{\sqrt{3}(12a_6^2c_6-2a_6b_6^2-18b_6c_6}{\sqrt{4a_6^3c_6-a_6^2b_6^2-18a_6b_6c_6+4b_6^3+27c_6^2}}-9b_6\right)a'_6\\
  a'_6&=&-8\frac{\rho_{ext}mGc_s^2}{\beta_0 p_{ext,0}\bar{R}}.
\end{eqnarray}
The derivative of $w$ is given by equation~\ref{e:wprime}, with $f'$ replaced by
\begin{equation}
  f'=-\frac{2f}{\sigma}\sigma'.
\end{equation}

%%%%%%%%%%%%%%%%%%%%%%%%%%%%%%%%%%%%%%%%%%%%%%%%%%%
\bibliographystyle{apj}
\bibliography{./references}
%%%%%%%%%%%%%%%%%%%%%%%%%%%%%%%%%%%%%%%%%%%%%%%%%%%

\end{document}